\documentclass[aps,prc,preprint,groupedaddress]{revtex4}
\usepackage{graphicx}
\def\Vec#1{\mbox{\boldmath $#1$}}

\begin{document}

%\preprint{}

\title{
Viability of nuclear $\alpha$-particle condensates \\
A reply to N.T. Zinner and A.S. Jensen, arXiv:nucl/th0712.1191
}


\author{Y.~\textsc{Funaki}, H.~\textsc{Horiuchi}$^{1,2}$, G.~\textsc{R\"opke}$^3$,
        P.~\textsc{Schuck}$^{4,5,6}$, A.~\textsc{Tohsaki}$^1$, T.~\textsc{Yamada}$^7$, and W.~\textsc{von Oertzen}$^{8,9}$}
%\thanks{}
\affiliation{Nishina Center for Accelerator-Based Science, The Institute of Physical and Chemical Research (RIKEN), Wako 351-0098, Japan}
\affiliation{$^1$Research Center for Nuclear Physics (RCNP), Osaka University, Osaka 567-0047, Japan}
\affiliation{$^2$International Institute for Advanced Studies, Kizugawa 619-0225, Japan}
\affiliation{$^3$Institut f\"ur Physik, Universit\"at Rostock, D-18051 Rostock, Germany}
\affiliation{$^4$Institut de Physique Nucl\'eaire, CNRS, UMR 8608, Orsay, F-91406, France}
\affiliation{$^5$Universit\'e Paris-Sud, Orsay, F-91505, France} 
\affiliation{$^6$Laboratoire de Physique et Mod\'elisation des Milieux Condens\'es, CNRS et Universit\'e
 Joseph Fourier, 25 Av.~des Martyrs, BP 166, F-38042 Grenoble Cedex 9, France}
\affiliation{$^7$Laboratory of Physics, Kanto Gakuin University, Yokohama 236-8501, Japan}
\affiliation{$^8$Hahn-Meitner-Institut Berlin, Glienicker Str.100, D-14109 Berlin, Germany}
\affiliation{$^9$Freie Universit\"at Berlin, Fachbereich Physik, Berlin, Germany}


\date{\today}

%\begin{abstract}
%@@@@@@@@@@@@@@@@@@@
%\end{abstract}

\maketitle

In a recent paper~\cite{zinner07}, arXiv:nucl/th0712.1191, 
 Zinner and Jensen (ZJ) expressed strong doubts about the concept
 of alpha-particle condensation in finite nuclei. 
In this article we give a reply which, essentially, is point by point
 (but not in the order).

\section{Definitions}

First let us define how we understand the concept of ``alpha-particle condensation in nuclei''. 
As explained in our previous work~\cite{schuck07}, the word ``condensation''
 is not to be understood in the macroscopic sense when talking about 
nuclei. 
It rather is to be seen in analogy to nuclear pairing, to nuclear 
 deformation and rotation, etc. 
Nuclear physicists became used to employ those macroscopic terms for things which
 are in reality only in a (slowly) fluctuating state. 
They well know this and it is only to be understood as a semantic short cut
 when they talk about ``nuclear superfluidity'', ``nuclear deformation'' and ``rotation'', etc. 
In reality e.g. the number of Cooper pairs in a nucleus is very limited and in no way
 one can consider this to be a macroscopic condensate. 
One only can say that an antisymmetrised product of Cooper pairs is a good approximation
 for certain nuclear states and phenomena which reflect pairing and superfluidity
 and that this product state goes continuously over into the macroscopic BCS state
 when the number of pairs is increased indefinitely. 
That is there is as much link between a macroscopic alpha-particle condensate
 and the product state of a few alphas in a nucleus as there is link between pairing
 in a nucleus with half a dozen of Cooper pairs and neutron superfluidity in a neutron star! 
Nobody will deny that such a strong corrspondence exists for the latter case. 
We then think that there is complete analogy between nuclear pairing, nuclear deformation,
 etc., and nuclear alpha-particle condensation, with the only difference that
 alpha-particle condensation has only recently been suggested as a new nuclear state. 
As is the case for pairing, this new nuclear property reflects, in a finite system,
 the state which it would acquire in a corresponding infinite system. 
For infinite nuclear matter alpha particle condensation has recently been suggested
 from a theoretical investigation to exist at low densities~\cite{roepke98,beyer00}. 
A quartet phase at low density was also found by a QMC solution of a $1D$ Hubbard model
 with four different fermions~\cite{quartet}. 
 We, therefore, define a state of condensed $n\alpha$'s, 
 if in a nuclear state the latter forms in good approximation a bosonic product state. So far in what concerns definitions.

\section{$\alpha$-particle density matrix}

One of the main points in the paper by ZJ is that they contest the uniqueness
 of the definition of our alpha-particle density matrix whose eigenvalues show
 that e.g. in the Hoyle state the three alpha particles form to nearly 75~\%
 a bosonic product state with the bosons all in the identical $0S$ state.
We have recently published a longer paper on this subject on arXiv~\cite{yamada08}
 and only repeat here the main conclusions. 
ZJ base their arguments on the fact that for self-bound systems like alpha-particles in nuclei,
 one necessarily has to define a density matrix corresponding to the intrinsic system
 where the center of mass coordinate has been removed. 
This question has recently been debated with respect to Bose-Einstein condensation
 of cold atoms~\cite{wilkin98,pethick00}.
N.K. Wilkin et al.~\cite{wilkin98} found that a BEC which rotates with
 its c.o.m.~in a trap potential but stays with its intrinsic state in the ground state,
 i.e.~no internal excitations are present, exhibits a so-called fragmented condensate,
 that is there are several eigenvalues of the single particle density matrix
 which show occupancies of the order of the total number of particles. 
This is to be contrasted with the situation of a uniform system where ALL condensed particles
 sit in the lowest momentum state $k=0$. 
In a subsequent paper Pethick and Pitaevskii (PP)~\cite{pethick00} argued
 that on physical grounds the situation of condensed particles should not be different
 in a uniform system from a Bose-system in a trap when the intrinsic system is not excited
 and that for that one has to work with a suitably defined density matrix of
 the ``internal'' system. 
Their internal density matrix is defined with ``internal'' coordinates
 $\Vec{q}_i = \Vec{r}_i - \Vec{R}$ where $\Vec{R}$ is the total c.o.m.~coordinate. 
Our study in Ref.~\cite{yamada08} shows, however, that with this so-defined internal
 density matrix one again obtains a fragmented condensate what is contrary to the initial
 claim and objective of PP. 
It turns out that the outcome of the study strongly depends on the definition of
 the internal coordinates: the coordinates chosen by PP are not orthogonal,
 this being the reason for the occurrence of a fragmented condensate. 
In choosing Jacobi coordinates which are orthogonal, we could show that bosons
 in a harmonic trap which form an ideal condensate in the laboratory frame,
 i.e.~all particles in the lowest 0S orbit, remain an ideal, i.e.~non-fragmented condensate, 
 once the c.o.m.~coordinate has been removed, that is internally. 
This, in agreement with the original objective of PP, seems to us the correct physical situation~\cite{pita}. 
In addition we could show that the internal density matrix defined with non-orthogonal
 coordinates leads to a fragmented condensate even in the macroscopic limit~\cite{yamada08}. 
At this point we should mention that in previous publications on alpha-particle condensation
 always the internal density matrix was defined with the Jacobi
 coordinates~\cite{matsumura04,yamada05,nupecc,funaki08}. 
We, therefore, conclude on this point that our previous statement that the Hoyle state
 in $^{12}$C is to nearly 75~\% a product state of three alpha particles condensed
 into an identical $0S$-orbit is {\it unambiguous}~\cite{matsumura04,yamada05,nupecc}. 
Similarly we recently have found in an extended investigation of $^{16}$O
 that the sixth $0^+$ state at 15.1 MeV also is a strong candidate to be
 of alpha-particle condensation nature with over 60~\% of the alpha-particles
 condensed~\cite{funaki08}. Therefore, those states fulfill our criterion of $\alpha$-particle condensation. 
At the same time, this brings to fall the main argument of ZJ which, initially, anyway was based on an erroneous formula~\cite{formula_ZJ}.

In the light of this finding, we would like to discuss again the content 
 of the THSR alpha-particle condensate wave function~\cite{thsr}. 
This wave function is given by
\begin{equation}
\Phi_{n\alpha}(B,b) = {\cal A} \left\{ \prod_{i=1}^{n} \exp \left(-\frac{2}{B^2}{\Vec{X}_i}^2 \right)\phi_{\alpha_i} \right\} \label{eq:1}
\end{equation}
with $\Vec{X}_i$ the coordinates of the c.o.m. motion of the $\alpha$-particles, and, e.g. 
\begin{equation}
\phi_{\alpha_1}({\Vec r}_1,{\Vec r}_2,{\Vec r}_3,{\Vec r}_4) = \exp\left(-\frac{1}{8b^2}\sum_{i<j=1}^4(\Vec{r}_i-\Vec{r}_j)^2\right).
\end{equation}
It is very important to remark, as is explained in Ref.~\cite{thsr}, that  this condensate wave function contains two 
 limits exactly. 
On the one hand, for $B=b$ we have a pure HO Slater determinant because the 
 antisymmetriser generates out of the product of simple Gaussians all 
 higher nodal wave functions of the HO. 
On the other hand, for $B\gg b$ the THSR wave function tends to a pure product state of alpha-particles, i.e. a mean field wave function,
 since in this case the antisymmetriser can be neglected. 
Indeed $B$ triggers the extension of the nucleus, i.e.~its average density. 
For alpha particles kept at their free space size (small $b$), the alpha-particles
 are then for large $B$-values far apart from one another and do not
 feel any action from the Pauli principle. 
The question is then whether, e.g. for the Hoyle state, the above wave function
 is closer to a Slater or to alpha-product state. 
Precisely this question is answered by the above discussed eigenvalues of 
 the density matrix. In this respect it is important to point out that in the 
calculation of the afore-mentioned density matrix always the total c.o.m. 
motion has been split off in the wave function of Eq.~(\ref{eq:1}) and that for the 
remaining relative c.o.m. 
coordinates the Jacobi ones have been used, as is clearly explained in \cite{matsumura04,yamada05}. In Refs.~\cite{matsumura04,yamada05} it has been shown, as explained,
 that the alphas occupy to over 70~\% the $0S$-orbit.
Therefore, the Hoyle state is in good approximation a product of
 three alpha particles, that is a condensate.

\section{Decay properties}

This brings us to a further critics of ZJ where it is claimed that
 besides the Hoyle state in $^{12}$C, no heavier self-conjugate nuclei
 can show analogous alpha-particle structure. 
The argument is based on the fact that the alpha-particle condensate states
 occur near the alpha-particle disintegration threshold which rapidly grows
 in energy and, thus, the level density in which such a condensate state
 is embedded raises enormously. 
For example the alpha-disintegration threshold in $^{12}$C is at 7.24 MeV
 and in $^{16}$O it is already at 14.4 MeV. 
Under ordinary circumstances this could mean that the alpha-particle condensate
 state in $^{16}$O, which we suppose to be the well known $0^+$ state at 15.1 MeV, has a very short life time and
 ZJ make a Fermi gas estimate in this respect. However, on the one hand
 it is a fact that the supposed $^{16}$O ``Hoyle''-state at 15.1 MeV
 has experimentally, for such a high excitation energy, a startling long 
lifetime (decay width 160 keV!) and on the other 
 hand it is easily understandable that such an exotic configuration as four 
 alpha-particles moving almost independently within the common Coulomb barrier, 
 has great difficulties to decay into states lower in energy which all have 
 very different configurations! 
How else one could explain such a long life time of a state at 15.1 MeV
 excitation energy? 
It is precisely one of the strong indications of alpha-particle condensation
 that the state should be unusually long lived! 
It is furthermore well known that the Hoyle state cannot be explained even
 with the most advanced shell model calculations. 
Its energy comes at 2-3 times its experimental value~\cite{nocore}.
This is a clear indication that shell model configurations 
 only couple extremely weakly to alpha condensate states. 
One can argue that many of the states in $^{16}$O
 below 15.1 MeV are of shell model type. 
There are also alpha-$^{12}$C configurations but since $^{12}$C also has shell
 model configuration, it again is difficult for the four alpha condensate
 state to decay into. 
This brings to fall a further argument of ZJ.

\section{Localisation}

Another critics of ZJ is that they say that a state of localized 
 alpha-particles can equally well describe the Hoyle state and they cite for 
 that the work of Chernykh et al.~\cite{chernykh07}. 
This again is a strong misconception. 
In the work of Chernykh et al. about 55 configurations are superposed. In our opinion these configurations mostly serve to {\it delocalise} the $\alpha$ particles~\cite{feldmeier}.

%~\cite{chernykh07} which just serve to mock up our single component wave function of the alpha-particle condensate character. Of course, as always, in choosing a basis which is not quite adapted to the physical situation, one still can recover the correct result in going to a large basis set! Does one learn much about the physics? In this context it should be mentioned that even for the description of $^{8}$Be, Feldmeier et al. have to use a quite extended basis to obtain a correct description~\cite{feldmeier}, whereas with the THSR wave function, this is almost a trivial task~\cite{funaki_8be}.

\section{The quantality condition}

The next point of ZJ is the least understandable. 
They claim on grounds of the so-called Mottelson ``quantality'' condition
 that a mean field description of freely moving alpha particles cannot be applied. 
Since our wave function is a prototype of a meanfield ansatz which leads, without
 any free parameter, to correct results for almost all measured quantities of the Hoyle state,
 this statement of ZJ can only be totally fallacious. 
On the other hand, the Gross-Pitaevskii equation was applied  
 to study dilute multi alpha-particle condensation in nuclei,
 in which we used a renormalized effective $\alpha\alpha$ potential~\cite{yamada04} (as always for a mean field).
This potential, of course, well fulfills the quantality condition.  
Also, using the energies of the mentioned resonances in $^{8}$Be and 
 $^{12}$C$^*$ and calculating the deBroglie wave length, we do find that the latter is 
 larger than the nuclear radius (see also~\cite{matsumura04} for same conclusion), a situation similar to the 
 pairing concept of neutrons in heavy nuclei. 
The alpha-clusters are in a condensed state, a fact 
 which probably has been observed experimentally by the emission of
 $^{12}$C$^*$~\cite{kokalova05} from excited $^{52}$Fe. Again on this point ZJ are advancing erroneous statements.

\section{Similarity of $\alpha$-particle condensates with varying particle number}

\begin{figure}[htbp]
\begin{center}
\includegraphics[scale=0.7]{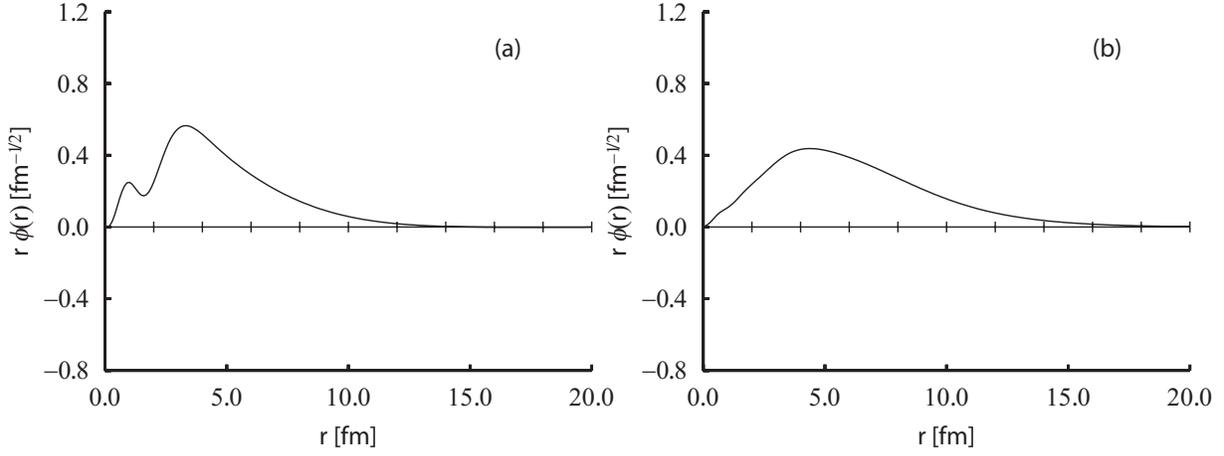}
\caption{Radial parts of the single-$\alpha$ $S$ orbits, (a) of the Hoyle state ($^{12}$C) and (b) of the $0_6^+$ state in $^{16}$O.}\label{fig:1}
\end{center}
\end{figure}

\begin{figure}[htbp]
\begin{center}
\includegraphics[scale=0.7]{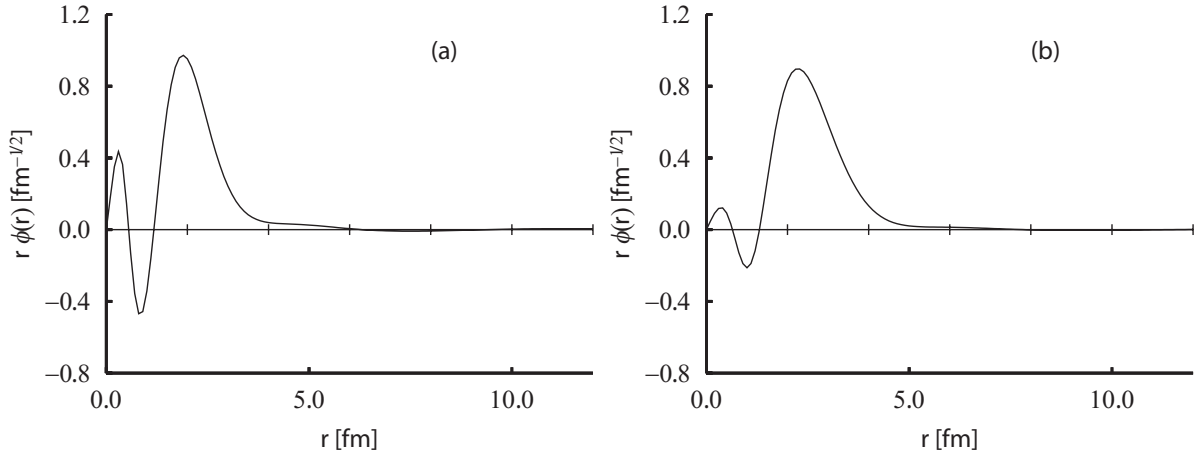}
\caption{Radial parts of the single-$\alpha$ $S$ orbits of the ground states, (a) in $^{12}$C and (b) in $^{16}$O.}\label{fig:2}
\end{center}
\end{figure}

In Fig.~\ref{fig:1} we show, side by side, radial parts of the single-$\alpha$ $S$ orbits (for definition, see Refs.~\cite{yamada05,funaki08,yamada08}) of the Hoyle state ($^{12}$C) and 
 the $0_6^+$ state in $^{16}$O~\cite{comments_fig}. 
We see an almost identical shape! 
Of course, the extension is slightly different because of the smallness of the system.
The nodeless character of the wave function is very pronounced and only some oscillations
 with small amplitude are present in $^{12}$C, reflecting a weak influence of the Pauli principle between the $\alpha$'s! 
On the contrary, we show in Fig.~\ref{fig:2} radial parts of the single-$\alpha$ $S$ orbits of the ground states
 in $^{12}$C~\cite{yamada05} and $^{16}$O~\cite{funaki08,comments_fig}. 
Due to its much reduced radius the ``$\alpha$-like'' clusters strongly overlap, producing strong amplitude
 oscillations which take care of antisymmetrisation between clusters. 
Again this example very impressively demonstrates the condensate nature of the Hoyle state and
 the $0_6^+$ state in $^{16}$O.
This result is much in contrast with the fact that ZJ announced the similarity criterion for $\alpha$-particles
 being very difficult to be fulfilled in finite systems with only a few bosons.

\section{Conclusion}

We have shown that the arguments of ZJ against the existence of 
 alpha-particle condensed states in self-conjugate nuclei are without 
 foundation. 
For instance we could very clearly demonstrate that their strongest 
 argument concerning the ambiguity of the eigenvalues of the density matrix is 
 false~\cite{yamada08,formula_ZJ}. We also could demonstrate the similarity of condensates
 with different number of $\alpha$-particles,
 another convincing argument in favor of the condensate aspect.


\begin{thebibliography}{99}

\bibitem{zinner07}
 N.T. Zinner and A.S. Jensen, arXiv:nucl/th0712.1191v1,v2,v3.

\bibitem{schuck07}
 For example, P. Schuck, Y. Funaki, H. Horiuchi, G. R\"opke, P. Schuck,
 A. Thosaki, and T. Yamada, Prog. Part. Nucl. Phys. 59, 285 (2007).

\bibitem{roepke98}
 G. R\"opke, A. Schnell, P. Schuck, and P. Nozieres, Phys. Rev. Lett. {\bf 80}, 3177 (1998).

\bibitem{beyer00}
 M. Beyer, S.A. Sofianes, C. Kuhrts, G. R\"opke, and P. Schuck, Phys. Lett. {\bf 448B}, 247, (2000).
 
\bibitem{quartet}
B. Doucot, J. Vidal, Phys. Rev. Lett. {\bf 88}, 227005 (2002); S. Capponi, G. Roux, P. Lecheminant, P. Azaria, E. Boulat, S.R. White, Phys. Rev. A {\bf 77}, 013624 (2007);  P. Lecheminant, P. Azaria, E. Boulat, arXiv:0711.4731. 

\bibitem{yamada08}
 T. Yamada, Y. Funaki, H. Horiuchi, G. R\"opke, P. Schuck, and A. Tohsaki, arXiv:cond/mat0804.1672.

\bibitem{wilkin98}
 N.K. Wilkin, J.M.F. Gunn, and R.A. Smith, Phys. Rev. Lett. {\bf 80}, 2265 (1998).

\bibitem{pethick00}
 C.J. Pethick and L.P. Pitaevskii, Phys. Rev. A {\bf 62}, 033609 (2000).

\bibitem{pita}
L.P. Pitaevskii, private communication.

\bibitem{matsumura04}
 H. Matsumura and Y. Suzuki, Nucl. Phys. A {\bf 739}, 238 (2004).

\bibitem{yamada05}
 T. Yamada and P. Schuck, Eur. Phys. J. A. {\bf 26}, 185 (2005).

\bibitem{nupecc}
 Y. Funaki, H. Horiuchi, G. R\"opke, P. Schuck, A. Tohsaki, and  T. Yamada,  Nucl. Phys. News {\bf 17} No.~4, 11 (2008).

\bibitem{funaki08}
 Y. Funaki, T. Yamada, H. Horiuchi, G. R\"opke, P. Schuck, and A. Tohsaki, arXiv:nucl/th0802.3246.

\bibitem{comments_fig}
The radial parts of the single-$\alpha$ $S$ orbits are normalised so as to satisfy $\int (r\phi(r))^2 dr =1$. That is why these are different by a factor $\sqrt{4\pi}$ from the orbits shown in Fig. 2 of Ref.~\cite{funaki08}, where $r\phi(r)$ is normalised as $4\pi \int (r\phi(r))^2 dr =1$.

\bibitem{formula_ZJ}
One should mention here that the original formula, Eq. (9) of ZJ~\cite{zinner07}
 is wrong (v1 and v2) and consequently the conclusions thereof also. 
After we published the right formula, ZJ also corrected their formula
 (with no citation of our work!), arXiv:nucl/th0712.1191v3, but left their conclusions unchanged!

\bibitem{thsr}
 A.~Tohsaki, H.~Horiuchi, P.~Schuck and G.~R\"opke, Phys. Rev. Let. {\bf 87}, 192501 (2001).

\bibitem{nocore}
 B. R. Barrett, B. Mihaila, S. C. Pieper, and R. B. Wiringa, Nucl. Phys. News, {\bf 13}, 17 (2003).

\bibitem{chernykh07}
 M. Chernykh, H. Feldmeier, T. Neff, P. von Neumann-Cosel, and A. Richter, Phys. Rev. Lett. {\bf 98}, 032501 (2007). 

\bibitem{feldmeier}
 H. Feldmeier and T. Neff, private communication.
%\bibitem{funaki_8be}
%Y. Funaki, H. Horiuchi, A. Tohsaki, P. Schuck and G. R\"opke, Prog. Theor. Phys. {\bf 108}, 297 (2002).

\bibitem{yamada04}
 T. Yamada, P. Schuck, Phys. Rev. C {\bf 69}, 024309 (2004).

\bibitem{kokalova05}
 Tz. Kokalova et al., Eur. Phys. J. A {\bf 23}, 19 (2005).

\end{thebibliography}
\end{document}